# Anti-vortex state in cross-like nanomagnets

V.L.Mironov, O.L.Ermolaeva, S.A.Gusev, A.Yu.Klimov, V.V.Rogov,

B.A.Gribkov, A.A.Fraerman, O.G.Udalov

Institute for Physics of Microstructures RAS, Nizhny Novgorod, 603950, GSP-105, Russia

R. Marsh, C. Checkley, R. Shaikhaidarov V.T. Petrashov

Department of Physics, Royal Holloway, University of London, Egham, Surrey TW20 0EX, United Kingdom

Electronic mail: mironov@ipm.sci-nnov.ru

We report on results of computer micromodelling of anti-vortex states in asymmetrical cross-like ferromagnetic nanostructures and their practical realization. The arrays of cobalt crosses with 1 μm branches, 100 nm widths of the branches and 40 nm thicknesses were fabricated using e-beam lithography and ion etching. Each branch of the cross was tapered at one end and bulbous at the other. The stable formation of anti-vortex magnetic states in these nanostructures during magnetization reversal was demonstrated experimentally using magnetic force microscopy.

## I. Introduction

Technological applications as well as fundamental investigations of magnetic nanostructures, the so-called nanomagnets, require well-controlled magnetic states. Examples are the single-domain nanomagnets that are very promising candidates for high-density hard disk data storage and integrated magnetoelectronic devices such as nonvolatile magnetic memory [1], nanomagnets with non-collinear magnetization which show a novel type of "helical" triplet superconductivity [2] and non-coplanar nanomagnets for that novel transport properties are expected [3-5]. One of the controls of the magnetic states at the nanometer scale is the size and geometry of nanoelements. Very small nanomagnets at the 10 nm scale are expected to behave as single giant spins since competition between magnetostatic energy and quantum mechanical exchange energy entirely suppresses magnetic domain formation. A range of interesting phenomena take place in the intermediate ("mesoscopic") range of 10-1000 nm, where electron-beam lithography allows for "magnetic nano-engineering" with well defined geometry of nanostructures. Highlights are the realization of vortex states in ferromagnetic nanodisks [1] and spiral (helical) states in laterally confined magnetic multilayers [6] and prediction of a novel type of magnetic walls in nanoconstrictions [7].

In this paper we report on a theoretical and experimental study of another fundamental magnetization structure, the anti-vortex [8-10], which is a topological counterpart of a magnetic vortex. Besides being a remarkable magnetic structure the anti-vortex is expected to show unusual transport properties in an applied external magnetic field, namely a spectacular new phenomenon, the so-called "topological" Hall effect [11].

Unlike a vortex, the realization of an anti-vortex, i.e. preparation of a nanostructure that contains only a single anti-vortex, is a challenging task. The anti-vortex possesses a "magnetic charge" resulting in additional, in comparison with a vortex, magnetostatic energy. A few years ago, an anti-vortex structure was created at a cross junction of four connected rings [10] with vortex distribution in each of the rings; however this complicated design does not allow for transport measurements, especially in an external magnetic field. In this work a single anti-vortex state has been realized in asymmetric cross-like nanoelements (nanomagnets) suitable for Hall effect measurements. The anti-vortex state is prepared by means of a specific procedure of magnetization reversal stimulated by the shape asymmetry.

We start with consideration of the distribution of magnetization in sufficiently thin ferromagnetic disks. In this case the magnetization depends only on the in-plane coordinates $\vec{M}(\vec{\rho})$, where $\vec{\rho}$ is the in-plane radius vector. In the absence of magneto-crystalline anisotropy and an external magnetic field the symmetric inhomogeneous distribution of magnetization in a circular disk can be represented in the following generalized equations [12]

$$\begin{aligned} M_x &= Sin\,\theta(\rho)\ Cos(\nu\varphi+\varphi_0), \\ M_y &= Sin\,\theta(\rho)\ Sin(\nu\varphi+\varphi_0), \\ M_z &= Cos\,\theta(\rho), \end{aligned} \quad (1)$$

where $\rho$ is the modulus of the radius vector $\vec{\rho}$; $\theta$ and $\varphi$ are polar and azimuth angles, respectively; $\varphi_0$ is an arbitrary phase shift; $\nu = 0, \pm 1, \pm 2 ...$ is the so-called "winding number". The topology of the magnetic state (1) is determined by the phase shift $\varphi_0$, direction of core magnetization ($\cos\theta(\rho=0) = p$) and winding number $\nu$. For a disk of sufficiently small radius and thickness the ground state is the uniform single-domain state ($p = \nu = 0$). In general non-uniform distributions can be realized with a magnetic structure determined by these parameters with polar angle independent of the phase shift $\varphi_0$ and the sign of winding number. For $\nu = 1, \varphi_0 = \pm\pi/2$ we have the well-known vortex states, where the sign of the phase shift $\varphi_0$ corresponds to clock-wise and counter clock-wise vortexes (fig. 1a). For $\nu = 1, \varphi_0 = 0, \pi$ we have a hedgehog-like distributions (fig. 1b). The set of parameters $\nu = -1, \varphi_0 = 0, \pi$ corresponds to the anti-vortex states (fig. 1c). In the last two cases we have a mismatch between the distribution of magnetization and the boundary of the particle, which leads to an increase of the energy of the system. For hedgehog-like states this excess in energy may be compensated by interaction of magnetic moments with the inhomogeneous external magnetic field induced by a magnetic force microscope (MFM) tip [13]. A decrease in the energy of an anti-vortex distribution can be realized by transformation of the shape of magnetic particle from circular to cross-like.

The density of effective "magnetic charge" for the anti-vortex is

$$div\,\vec{M} \sim \mp\frac{1}{\rho}Cos2\varphi,\ \rho \gg l. \quad (2)$$

The sign $\mp$ in this formula corresponds to a phase shift $\varphi_0 = 0, \pi$. The parameter $l$ is the anti-vortex core radius and for typical transition metals this value is about 20 nm.

The distribution of "magnetic charge" induces a magnetic stray field and as a result we have an additional contribution in the energy of the anti-vortex, which is proportional to the particle volume. So, the total energy (= the sum of exchange and magnetostatic energies) of quasi uniform distribution in cross-like structures (fig. 2a) is smaller than the energy of a stable anti-vortex (fig. 2b) making spontaneous realization of an anti-vortex state a hard problem. Moreover, a decrease in the aspect ratio $g=a/b$ of the elements leads to the formation of a vortex distribution (fig. 2c) that narrows the range of dimensions for the realization of the anti-vortex structure.

An additional contribution to the Hall current mentioned above, the "topological" Hall current arises in the form:

$$\vec{j} = \left[ \vec{E} \times \vec{B}^{eff} \right],$$

$$B_x^{eff} = \alpha \left( \vec{M}, \left[ \frac{\partial \vec{M}}{\partial y} \times \frac{\partial \vec{M}}{\partial z} \right] \right),$$

$$B_y^{eff} = \alpha \left( \vec{M}, \left[ \frac{\partial \vec{M}}{\partial z} \times \frac{\partial \vec{M}}{\partial x} \right] \right), \quad (3)$$

$$B_z^{eff} = \alpha \left( \vec{M}, \left[ \frac{\partial \vec{M}}{\partial x} \times \frac{\partial \vec{M}}{\partial y} \right] \right).$$

The winding number $\nu$ has a direct impact on the magnetization dynamics and transport properties of inhomogeneous ferromagnets [14, 15]. Combining formulas (1) and (3) we obtain the following expression for the effective magnetic field $B^{eff}$

$$B_z^{eff}(\rho) \sim \alpha \nu \frac{1}{\rho} \frac{\partial M_z}{\partial \rho}; \quad B_x^{eff} = B_y^{eff} = 0, \quad (4)$$

where the constant $\alpha$ is of the order of flux quantum $\Phi_0 \sim 10^{-7}$ Oe×cm$^2$ [4,15]. It can be seen that the sign of "topological" Hall current is determined by the winding number and the core polarization ($p$) and does not depend on the phase shift $\varphi_0$. Observation of the topological Hall effect in ferromagnetic nanostructures and control of the sign of the effect through control of the winding number is a very attractive possibility yet to be realized.

In section II of the paper we present the results of computer micromagnetic modeling of the magnetization reversal process in an asymmetrical cross. Section III is devoted to the description of experimental results, which illustrate practical realization of anti-vortex states.

**II. Computer micromagnetic modeling**

The micromagnetic modeling of magnetic states was performed using software [16] based on solutions to the Landau-Lifshitz-Gilbert (LLG) equation [17]. In the model we have taken into account the dependence of coercivity of magnetic elements on their shape [18] and considered an asymmetrical cross shown in fig. 3. The magnetic structure in the cross can be manipulated by an external magnetic field directed along one of the diagonals of the cross as follows (fig. 3):

a)  After magnetization in a sufficiently strong (~1 kOe) magnetic field we obtained a quasi uniform distribution (fig. 3a).

b)  Then, under the action of a certain critical magnetic field $H$ with magnitude $H_1 < H < H_2$ ($H_1$ and $H_2$ are the coercivity of bulbous and tapered ends respectively) oriented in the opposite direction we observed at the first stage the nucleation of vortexes in the bulbous segments (fig. 3b).

c)  At the second stage the nucleation of an anti-vortex core with extra two vortexes near the centre of the cross (fig. 3c) was observed.

d)  At the final stage the annihilation of the vortexes and formation of the anti-vortex distribution of magnetization in the cross were observed (fig. 3d).

The magnetization in the tapered elements was kept in the initial uniform state at all stages. Thus after the action of the reversed magnetic field $H_1 < H < H_2$ we obtained the anti-vortex distribution.

## III. Experimental results

The arrays of Co crosses were fabricated using a negative e-beam lithography process. A film of Co with thickness 40 nm was deposited onto a Si substrate using magnetron sputtering and covered with 100 nm thick positive UV photoresist (FP-9120), which is based on phenol-aldehyde resin. The exposure was done using the ELPHY PLUS system based on the scanning electron microscope SUPRA 50VP. The magnetic states in the Co crosses were studied using a multimode scanning probe microscope "Solver-HV". The scanning probes were cobalt-coated with a thickness of 30 nm. Before measurements the tips were magnetized along the symmetry axes (Z) in 10 kOe external magnetic field. The MFM measurements were performed in the non contact constant height mode. The phase shift of cantilever oscillations under the gradient of the sample magnetic field was registered to obtain the MFM contrast.

The first set of measured samples was consisted of symmetric crosses. Their magnetic structure depended on the aspect ratio, as expected. Crosses with relatively small aspect ratio $g$ showed spontaneous quasi-uniform states. The crosses with larger aspect ratio $g$ could go into vortex states. Spontaneous quasi-uniform states with two types of symmetry, A and B were observed in the crosses with lateral size $a = 600$ nm and width of branch $b = 100$ nm (fig. 4 and fig. 5 respectively).

A characteristic feature of A and B uniform states is the quadrupole symmetry of the MFM contrast distribution (see fig. 4b,c and fig. 5b,c). On the other hand the expected MFM contrast corresponding to the anti-vortex distribution of magnetization (fig. 2b) is represented in fig 6.

For the crosses with lateral size $a = 600$ nm and width of branch $b = 200$ nm quasi-vortex states were registered. The experimental MFM image of the quasi-vortex state and corresponding model picture are represented in fig. 7.

To realize the anti-vortex state we fabricated asymmetrical crosses analogous to the elements considered in micromagnetic modeling (fig. 3). The SEM image of fabricated structure is represented in fig. 8. The lateral size of the crosses was $a = 1$ μm, the width of the branch was $b = 100$ nm and size of the bulb was 150 nm.

Fig. 9 shows the results of MFM investigation of asymmetric crosses. Magnetization in a strong (800 Oe) magnetic field directed along the diagonal of the cross (as shown in fig. 3a) results in quasi uniform magnetization (fig. 9a). A weak reversed magnetic field with magnitude of 250 Oe leads to the transition of one of the particle into anti-vortex state with characteristic quadrupole symmetry of MFM contrast (fig. 9b). An increase of the external magnetic field to 400 Oe leads to the creation of anti-vortexes in all crosses (fig. 9c). The observed dispersion of switching fields from 250 Oe to 400 Oe can be connected with the dispersion in particle coercive forces due to shape differences.

**IV. Conclusions**

We reported the results of micromagnetic model LLG simulations and MFM investigations of magnetic states in ferromagnetic cross-like nanomagnets. It was shown that shape control based on electron-beam lithography allows for practical "nano-engineering" of magnetic states in cross-like nanostructures. We demonstrated that symmetric Co nano-crosses can be put into spontaneous quasi-uniform and quasi-vortex states depending on their size and aspect ratio. The asymmetrical Co crosses where each branch is tapered at one end and bulbous at the other showed the stable formation of anti-vortex states during magnetization reversal. These structures with controllable quasi-uniform and anti-vortex states are very promising for the investigations of transport peculiarities and magneto-dynamical phenomena in inhomogeneous magnetic systems.

**Acknowledgements**

This work was supported by the Russian Foundation for Basic Research (project 08-02-01202), Russian Federal Educational Agency (contract № P 348 and № P 417), Russian Federation President's grant for support of young scientists (Grant № MK-4508.2009.2) and the UK EPSRC.

**FIG.1**

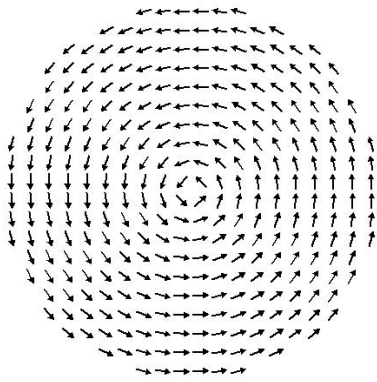 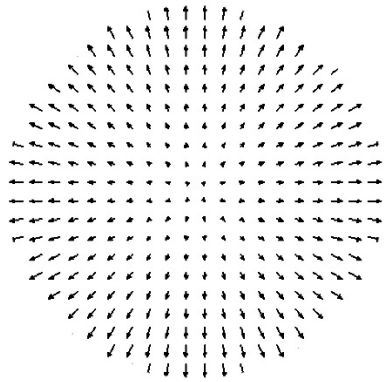 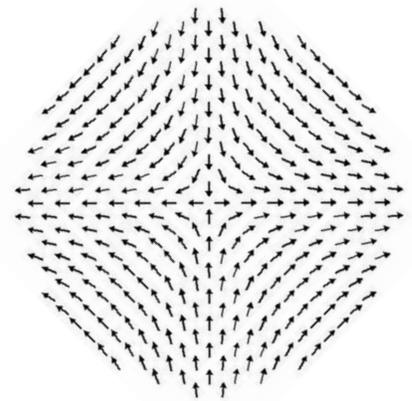

(a)  (b)  (c)

**FIG.2**

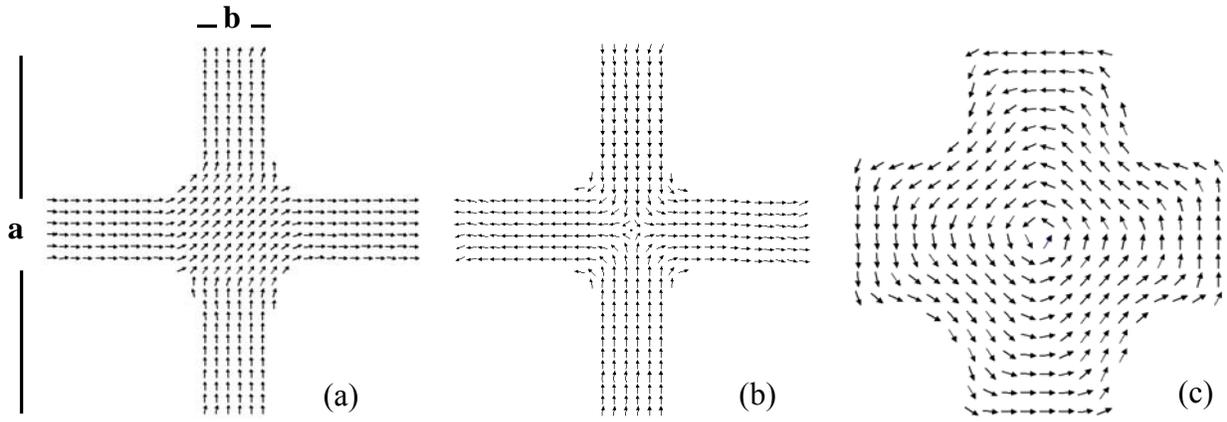

(a)   (b)   (c)

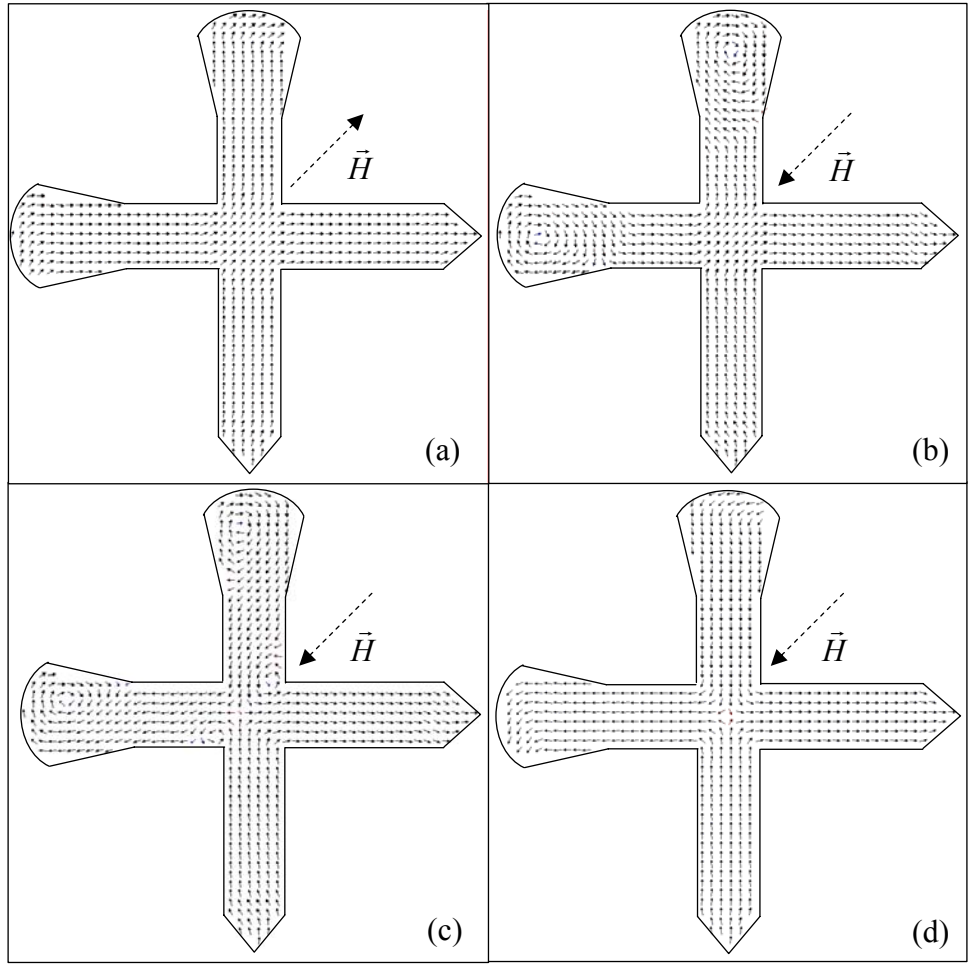

**FIG. 4**

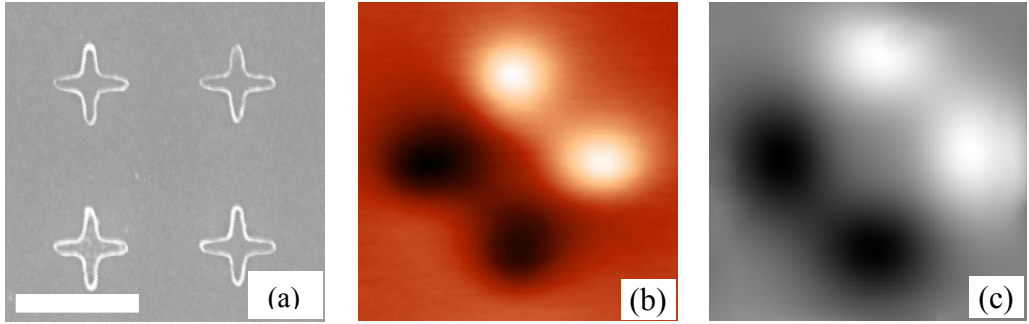

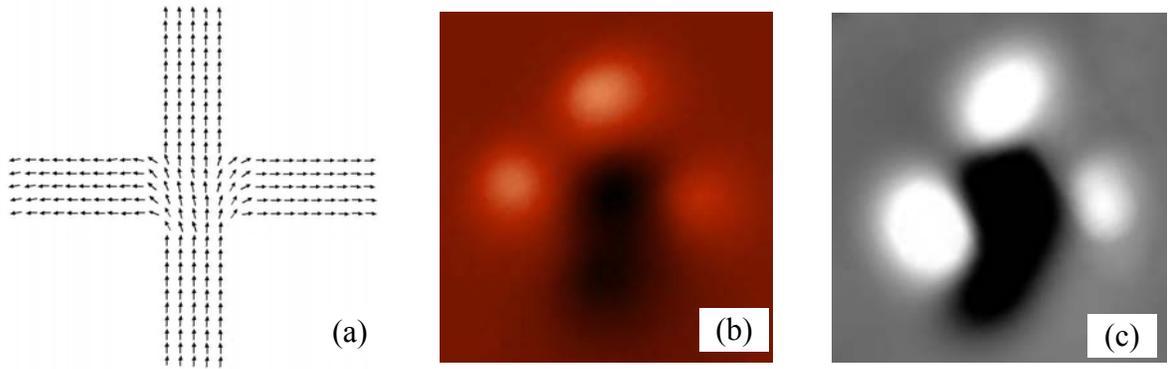

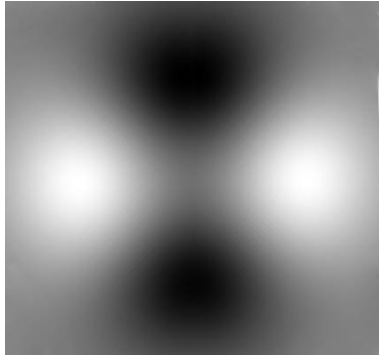

**FIG. 7**

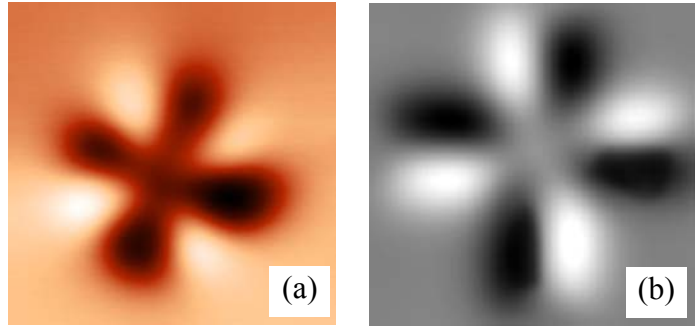

**FIG. 8**

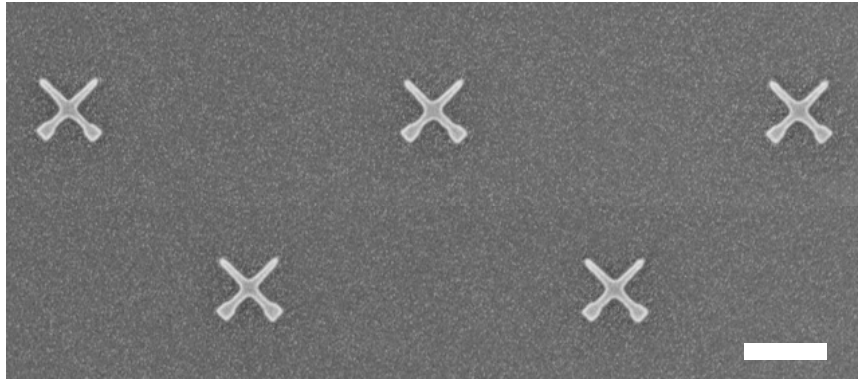

**FIG. 9**

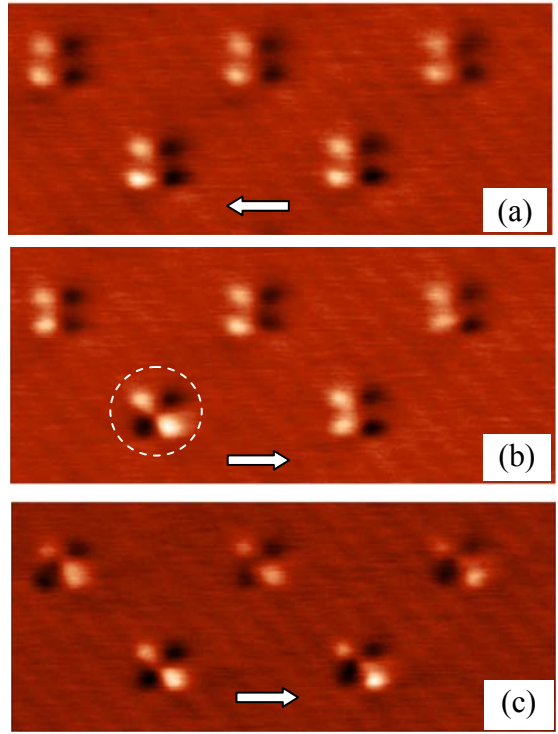

Fig. 1. The distributions of magnetization for the magnetic states with winding number equal to one. (a) is a vortex state $v=1, \varphi_0=+\pi/2$, (b) is a hedgehog-like state $v=1, \varphi_0=0$, (c) is an anti-vortex state $v=-1, \varphi_0=0$.

Fig. 2. The model distributions of magnetization in cross-like elements. (a) is for quasi uniform state, (b) is for anti-vortex state, (c) is for vortex state.

Fig. 3. The formation of an anti-vortex state during the re-magnetization of an asymmetrical cross in an external magnetic field. (a) is the initial distribution of magnetization; (b) is the nucleation of vortexes in the reversed external magnetic field; (c) is the nucleation of an anti-vortex core and formation of additional vortexes; (d) is the final anti-vortex state. The lateral size of the crosses $a$ was about 1 μm; the width of the segments $b$ was 100 nm and size of the bulb was 150 nm.

Fig. 4. Magnetic states in symmetrical crosses. (a) SEM image of an array of symmetric cross-like cobalt nanostructures. The scale bar is 1 μm. Experimental (b) and model (c) MFM contrast distributions corresponding to the A type of quasi-uniform state of magnetization (fig. 2a).

Fig. 5. (a) is the B type quasi-uniform state with three incoming and one outgoing magnetization vectors (simulation). The experimental (b) and model (c) MFM contrast distribution corresponding to the B type of quasi-uniform state of magnetization.

Fig. 6. The model MFM contrast distribution corresponding to the anti-vortex state (fig. 2b) of magnetization.

Fig. 7. The experimental (a) and model (b) MFM contrast distributions corresponding to the quasi-vortex distribution of magnetization (fig. 2a).

Fig. 8. SEM image of asymmetric Co nano-crosses. The white scale bar is 1 μm.

Fig. 9. Transformation of magnetic states in asymmetric Co crosses from quasi uniform states to anti-vortex states in an external magnetic field. (a) is MFM image of the initial quasi-uniform state. (b) is MFM image of the same array of crosses in a weak (250 Oe) reversed magnetic field. The anti-vortex state is indicated by a white circle. (c) is MFM image of the final anti-vortex states. The directions of the applied magnetic fields are indicated by white arrows. The frame size is 6 × 12 μm.